# Ultrafast charge transfer processes accompanying KLL Auger decay in aqueous KCl solution

D. Céolin[1], N. V. Kryzhevoi[2]*, Ch. Nicolas[1], W. Pokapanich[3], S. Choksakulporn[3], P. Songsiriritthigul[4,5], Th. Saisopa[4,5], Y. Rattanachai[6], Y. Utsumi[1], J. Palaudoux[7,8], G. Öhrwall[9], J. P. Rueff[1,7,8]

[1] Synchrotron SOLEIL, l'Orme des Merisiers, Saint-Aubin, F-91192 Gif-sur-Yvette Cedex, France
[2] Theoretical Chemistry, Institute of Physical Chemistry, Heidelberg University, Im Neuenheimer Feld 229, 69120 Heidelberg, Germany
[3] Faculty of Science, Nakhon Phanom University, Nakhon Phanom 48000 Thailand
[4] Synchrotron Light Research Institute, Nakhon Ratchasima 30000, Thailand
[5] NANOTEC-SUT Center of Excellence on Advanced Functional Nanomaterials and School of Physics, Suranaree University of Technology, Nakhon Ratchasima 30000, Thailand
[6] Department of Applied Physics, Faculty of Sciences and Liberal Arts, Rajamangala University of Technology Isan, Nakhon Ratchasima 30000, Thailand
[7] CNRS, UMR 7614, Laboratoire de Chimie Physique-Matière et Rayonnement, F-75005, Paris, France
[8] Sorbonne Universités, UPMC Université Paris 06, UMR 7614, Laboratoire de Chimie Physique-Matière et Rayonnement, F-75005, Paris, France
[9] MAX IV Laboratory, P.O. Box 118, SE-22100 Lund, Sweden

## Abstract

X-ray photoelectron spectroscopy (XPS) and KLL Auger spectra of aqueous KCl solution were measured for the $K^+$ and $Cl^-$ edges. While the XPS spectra of potassium and chloride have similar structures, both exhibiting only weak satellite structures near the main line, the Auger spectra of these isoelectronic ions differ dramatically. A very strong satellite peak was found in the $K^+$ KLL Auger spectrum at the low kinetic energy side of the $^1D$ state. Using equivalent core models and *ab initio* calculations this spectral structure was assigned to electron transfer processes from solvent water molecules to the solvated $K^+$ cation. Contrary to the potassium case, no extra peak was found in the KLL Auger spectrum of solvated $Cl^-$ indicating on a strong dependence of the underlying processes on ionic charge. The observed charge transfer processes are suggested to play an important role in charge redistribution following single and multiple core-hole creation in atomic and molecular systems placed into an environment.

Charge transfer (CT) processes and related phenomena are topics of wide relevance in chemistry, physics and biology. They are responsible for numerous important transformations in living organisms and are involved in fundamental steps describing, e.g., the photosynthesis [1] and respiration [2] mechanisms. The creation of a charge (or hole) in a DNA chain, by oxidation or photoionization processes, is quickly followed by its migration along portions of the molecular backbone leading to possible bonds breaking and irreversible damages [3]. Naturally, the use of such very fast elemental processes for technological purposes has attracted an extraordinary large amount of scientists whose research programs – both theoretical and experimental – focus on energy conversion based for instance on photovoltaic or opto-electronic devices [see e.g. ref. [4]].

CT may accompany core-hole creation in atoms and molecules which have neighbors, and the corresponding spectral signatures are manifested in XPS spectra as low-energy CT satellites. Particularly strong CT screening satellites were found in XPS spectra of weak chemisorption systems [5,6], in crystals [7,8], in weakly bound atomic [9] and microsolvated [10-12] clusters. As shown in the latter studies, the energy positions and intensities of CT satellites are sensitive to cluster geometries. Furthermore, the type and number of neighbors have strong effects on CT states [13]. CT processes from core-excited metal ions to solvent molecules may quench radiative relaxation processes as observed in fluorescence-yield spectra [14].

For elements from the first rows of the periodic table, electronic Auger decay is the main relaxation channel of core-hole states. Core-hole lifetime ranging usually in the femtosecond and sub-femtosecond timescales may serve as an internal reference clock. Using this reference and the relative intensities of spectral peaks, the core-hole clock method allows one to determine timescales of various processes competing to Auger decay, in particular timescales of ultrafast CT processes induced by inner-shell ionization [15,16]. For purely resonant core-excited states, the use of a photon bandwidth which is narrower than the core-hole lifetime broadening, makes a control over fast dynamical processes possible as demonstrated in Ref. [17].

Recently, the core-hole clock method was applied for determination of the timescale of intermolecular Coulombic decay (ICD) in core-ionized aqueous solutions [18]. Contrary to Auger decay, ICD is a non-local electronic decay process which involves environment. Originally, it was predicted for inner-valence ionized weakly bound systems [19] but can also operate after core ionization and efficiently compete with local Auger decay [20]. ICD leads to creation of two distributed electron vacancies, one of which is outside of the initially ionized moiety. Another non-local electronic decay process, electron transfer mediated decay (ETMD) [21], produces two holes entirely in the local environment of the initially ionized system. This process was also observed in aqueous solution [22]. It is believed that non-local electronic decay processes are responsible for charge redistribution occurring after ionization of a single and multiple core electrons in clusters by intense X-ray free-electron-laser pulses, and, thus, represent important mechanisms of nanoplasma formation [23].

In the present study we address another, hitherto nearly unexplored, mechanism of charge redistribution occurring after core ionization. In the first step of the corresponding process, the created core-hole state, typically $1s^{-1}$, undergoes KLL Auger decay in which two 2p deep core holes are produced. A sudden creation of two core vacancies induces even larger perturbation in the valence region compared to the initial core ionization leading to strong CT satellites in the Auger spectrum. Such satellites were already detected in Auger spectra of some solid state samples [24]. Here, we observe them for the first time in a weakly bound system, namely in KCl aqueous solution. Importantly, the corresponding CT processes involve solvent water molecules which donate their electrons to the solute and thus gets ionized in the end. We also note that in contrast to ICD peaks appearing at the high kinetic energy side of Auger peaks [18, 25], the CT peak observed in the present study has lower kinetic energy than the Auger peak. Thus, ICD and CT processes can be distinguished spectroscopically.

**Results:**

The 1s XPS spectra of Cl$^-$(aq) and K$^+$(aq) were recorded at hν=5 keV far enough from the 1s ionization thresholds to avoid post collision interaction effects (figure 1). For calibration of the core photoelectron lines we used the O1s photoelectron line of liquid water with the binding energy of 538.1 eV [26]. The 1s ionization potentials of Cl$^-$ and K$^+$ were found at 2825.4 eV and 3611.9 eV, respectively. The Lorentzian contributions extracted from of the main lines have the full width at half-maximum of 0.72 eV for K$^+$(aq) and 0.62 eV for Cl$^-$(aq), and compare well with the theoretical values obtained for bare potassium (0.74 eV) and chlorine (0.64 eV) atoms [27]. These values correspond to the 1s core-hole lifetimes of 0.9 and 1 fs, respectively. As seen from figure 1, each core level spectrum exhibits a week structure separated from the main line approximately by 4.5 eV in the case of chloride and by 6.0 eV in the case of potassium. Partly this structure can be attributed to inelastic scattering processes and the onset energy matches the band gap of liquid water quite well [28]. It may also contain discrete satellites states which are discussed below.

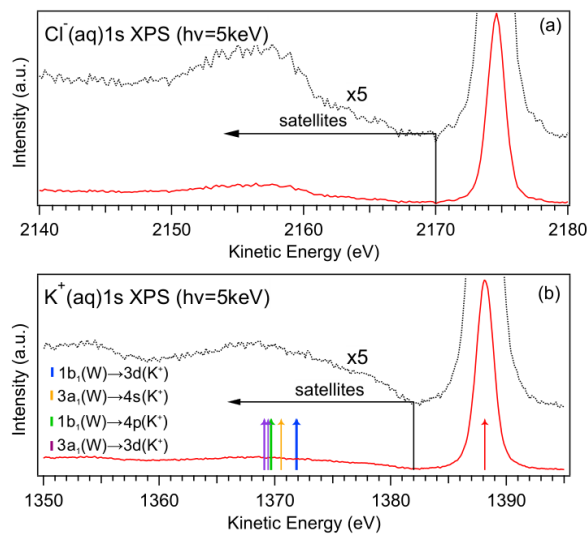

**Figure 1**: The 1s XPS spectra of Cl$^-$ (top) and K$^+$ (bottom) in aqueous KCl solution recorded at photon energy hν=5keV. A x5 zoom (black line) in each spectrum highlights the presence of satellite states. The arrows in the potassium spectrum indicate the positions of the main state and low-energy satellites in the computed spectrum of a K$^+$(H$_2$O) cluster. The characters of the satellites are represented in the legend.

The 1s core-ionized K$^+$ and Cl$^-$ ions are unstable and relax mostly via Auger decay processes. The KL$_{2,3}$L$_{2,3}$ Auger decay produces $2p^{-2}$ double core hole states represented by $^1S$, $^1D$ and $^3P$ terms, wherein Auger transitions giving rise to the $^3P$ term are forbidden from parity-conservation rules (cf. figure 2). Information on the energy positions of these lines for chlorine and potassium containing molecules is rather scarce, and to best of our knowledge is not available in the literature for solvated ions.

Cleff and Mehlhorn measured the Cl(KLL) Auger spectrum of CCl$_4$ using a 9keV electron beam and found the $^1D$ line at the kinetic energy of 2382.8eV±1eV [29]. This experimental result agrees well with earlier theoretical results predicting the kinetic energy positions of the $^1S$, $^1D$ and $^3P$ lines at 2370 eV, 2382 eV and 2391 eV, respectively [30]. Vayrynen et al. [31] measured the KLL Auger spectrum of chlorine in HCl and determined the positions of the $^1S$ and $^3P$ lines at -8.5eV and 6.4eV, respectively, relative to the position of the $^1D$ line at 2372.3 eV. The latter agrees with a measurement of Aitken et al. giving a value of 2371.98 eV [32]. The KLL Auger spectrum of aqueous $Cl^-$ has a very similar shape (see figure 2a). We found two strong peaks at the kinetic energies of 2372 and 2381 eV which were attributed to the $^1S$, and $^1D$ states, respectively, and a weaker structure at ~2388.5 eV assigned to the $^3P$ state.

For potassium, the literature is even poorer. Lohmann and Fritzsche [33] calculated the KLL Auger spectrum of atomic potassium and obtained the kinetic energy of 2959.82 eV for the $2p^4(^1D)4s$ state, 2950.49 eV for the $2p^4(^1S)4s$ state, and an energy window ranging from 2966.73 to 2969.62 eV for the $2p^4(^3P)4s$ state. The KLL Auger spectrum of aqueous $K^+$ has a remarkably different shape. We found two peaks and a weaker structure at the kinetic energies of 2957, 2967.4 and ~2976.3 eV, which were assigned to the $^1S$, $^1D$ and $^3P$ states, respectively. In addition, a very intense extra peak was discovered at the low kinetic energy side of the $^1D$ state, at 2963 eV, which is not present in the Auger spectrum of chloride. We note that neither the change in salt concentration from 0.5M to 1M, nor the decrease of the photon flux by a factor of 10 modified noticeably the shape of the potassium spectrum.

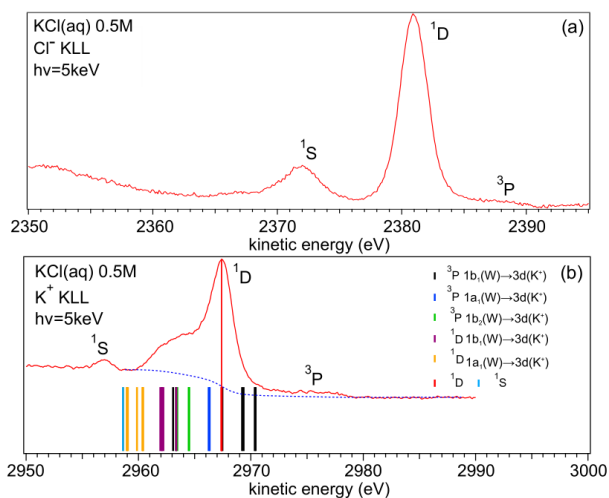

Figure 2: The KLL Auger spectra of $Cl^-$ (top) and $K^+$ (bottom) in aqueous KCl solution recorded at photon energy hv=5keV. In the potassium case, a Shirley background is shown by blue dotted curve. The vertical bars indicate the energy positions of the calculated singlet $2p^{-2}$ double core hole states in a $K^+(H_2O)$ cluster. The characters of these states are given in the legend.

## Discussion:

XPS spectra: Before we explain the KLL Auger spectrum of the aqueous potassium cation, let us first look again at the XPS spectra of both ions. Compared to bare anions whose K-shell photodetachment spectra have satellites of entirely local character, the spectra of solvated anions are more complex possessing energy loss features and non-local excitations, for example, to charge-transfer-to-solvent states [34]. Local and non-local satellites may appear in the same energy region near the main photoelectron line as demonstrated recently for microsolvated $Li^-$ [35]. Presumably, both local and non-local low-energy satellites are also present in the XPS spectrum of aqueous $Cl^-$ (figure 1a). The discrimination of these satellites is an interesting subject that is, however, beyond the scope of the present paper.

The XPS spectrum of aqueous $K^+$ (figure 1b) is less ambiguous. Indeed, according to our calculations the first satellite in the XPS spectrum of bare $K^+$ is separated from the main line by about 40 eV. Thus, the low-energy satellite structure seen in the spectrum of solvated $K^+$ must not be due to local excitations but rather have a non-local nature and originate due to the presence of solvent water molecules. To gain more insight we computed energies of several core-ionized states in microsolvated $K^+(H_2O)$ and $K^+(H_2O)_2$ clusters (see Computational methods). The results for the monohydrated cluster are shown in figure 1b as arrows (an energy shift of 1.87 eV was applied to adjust the main lines in the experimental and theoretical spectra). All low-energy satellites in the theoretical spectra have a CT character corresponding to excitations of valence electrons of water molecules to vacant orbitals of the potassium cation. The first ones are separated from the main line by about 16 eV and attributed to excitations of the water's $1b_1$ electron to an unoccupied 3d orbital of potassium, $1b_1(H_2O) \rightarrow 3d(K^+)$. These satellites are followed by the $3a_1(H_2O) \rightarrow 4s(K^+)$ and $1b_1(H_2O) \rightarrow 4p(K^+)$ CT states. In the dihydrated cluster, the energy separation between the main line and the first satellites increases up to 20.5 eV (not shown). The same trend was found also in the XPS spectra of microsolvated $Na^+$ [12].

As one can see, the energies of the CT satellites in the theoretical spectra agree well with the position of the maximum of the experimental satellite structure indicating that the latter has a CT origin. The respective CT processes are rather week though. This spectral feature has certainly contributions also due to inelastic scattering processes, in particular at the onset where our cluster models predict no states.

KLL Auger spectra: The extra peak seen in the $K^+$ KLL Auger spectrum, at ~3 eV below the $^1D$ state, cannot be due to Auger processes of core-ionized CT satellites since the latter are only weekly populated. Indeed, the intensity ratio of the main line and satellites in the XPS spectrum is ~14, while the intensity ratio of the $^1D$ line and the extra peak in the KLL spectrum is about 3 (after subtraction of the Shirley background shown in figure 2b by blue dotted curve). This peak cannot be explained alone by the inelastic scattering processes. The spectral contribution of these processes is expected to be rather week and similar to that in the XPS spectrum. The extra peak in the Auger spectrum has therefore a different origin.

Similar structures were found earlier in KLL Auger spectra of various potassium compounds and assigned to electron excitations from ligands to core-ionized potassium [24]. Since counterions are far separated from each other in the low-concentrated KCl aqueous solution studied here, chloride anions can hardly be involved in the above CT processes. On the other hand, as we already know, solvent water molecules can readily donate their electrons to neighboring core-ionized cations.

Assuming that water is involved in the CT processes, the difference between the energy of the main line, $E_{ML}$, and the energy of a CT satellite, $E_{CT}$, in the KLL Auger spectrum of potassium can be expressed as follows

$$\Delta E = E_{CT} - E_{ML} = E(K^{2+}_{cc}\ H_2O^+) - E(K^{3+}_{cc}\ H_2O) + RE, \qquad (1)$$

where the subscript c denotes a core hole, and RE is the Coulomb repulsion energy between a potassium dication and a water cation. Using the equivalent core model, we substitute the core-ionized potassium by its equivalent-core ion, Sc, and get

$$\Delta E = E_{CT} - E_{ML} = E(Sc^{2+}\ H_2O^+) - E(Sc^{3+}\ H_2O) + RE =$$
$$= -IP(Sc^{2+}) + IP(H_2O) + RE, \qquad (2)$$

where IP denotes outer-valence ionization potentials in solution. Contrary to the well-known IPs of liquid water (the lowest one is 11.16 eV [36]), those of aqueous $Sc^{2+}$ are not available since this dication does not exist in aqueous solutions. We shall use a value of 8.76 eV which is obtained by subtracting the solvation shift of aqueous $Ca^{2+}$ (~16 eV [37]) from the IP of gas-phase $Sc^{2+}$ (24.76 eV [38]). By substituting this value and the IP of liquid water into Eq. 2, and neglecting for the moment the repulsion energy RE (the Coulomb interaction is largely screened in solutions [23,26]), we arrive at $\Delta E$ of 2.4 eV. This energy difference fits well to the experimental observation.

Note, that this is a rather crude estimate. The Coulomb interaction is not completely screened in aqueous solutions and residual repulsion energy should be added to $\Delta E$. We also do not know yet whether $Sc^{2+}$ is in a ground state or in an excited one. Ionization potentials of excited states are apparently lower. In order to gain more insight into the CT processes accompanying the KLL Auger decay in aqueous potassium we computed the energies of the final singlet Auger states in the $K^+(H_2O)$ and $K^+(H_2O)_2$ clusters (triplet states do not acquire much intensity in the experimental spectrum). The low-lying states in the monohydrated cluster are shown in figure 2b as bars. We shifted them to adjust the energy positions of the $^1D$ states.

The first CT satellites in the theoretical Auger spectrum appear even at higher kinetic energies than the main $^1D$ state. These states are however related to the parent $^3P$ term and, thus, should not acquire much intensity. The first CT transitions in the presence of the $2p^{-2}(^1D)$ double-core hole are found at about 4 eV below the $^1D$ state, in well agreement with the experiment. They correspond to transfer of the water's $1b_1$ electron to an unoccupied 3d orbital of potassium, $1b_1(H_2O) \rightarrow 3d(K^+)$. These transitions are followed by the $3a_1(H_2O) \rightarrow 3d(K^+)$ ones. Note that electron transfer to the 4s and 4p orbitals of potassium requires higher energies and the corresponding CT states appear only at lower kinetic energies than the $^1S$ main state.

## Conclusions:

In conclusion, we performed the first high kinetic energy photoemission experiments on a KCl aqueous solution and discovered CT processes accompanying single and double core-hole creation in solvated $K^+$ which were not observed in aqueous media before. The respective CT occurs from solvent water molecules. Being not very evident in the XPS spectrum of $K^+(aq)$ because of the overlap with the energy-loss spectral feature, CT is well manifested in the KLL Auger spectrum giving rise to a pronounced extra peak. Large relaxation of valence electrons induced by creation of two 2p core holes in the KLL Auger decay seems to play a decisive role in the CT processes. Note that CT processes accompany neither the LMM Auger decay in $K^+(aq)$ and $Ca^{2+}(aq)$ [18,25] nor the KLL Auger decay in solvated $Na^+$ [37]. The corresponding Auger processes create outer-valence double vacancies and do not lead to strong electron reorganization in the valence shell. Interestingly, although the KLL Auger decay in $Cl^-(aq)$ also creates a double core hole, no indication of CT was found in the respective spectrum. The ionic charge seems to be an important parameter for CT processes. On the other hand, the ion-water distances are larger for $Cl^-$ than for $K^+$ (3.16 Å vs. 2.65 Å according to ref. [39]) and this should also affect CT. Further studies are needed to gain more insight here.

## Methods:

Experimental: Data were collected with the newly operational setup dedicated to measurements on liquids, specifically designed for the HAXPES station [40] of the GALAXIES beamline [41] located at the SOLEIL synchrotron facility, France. The details of this setup based on the development presented in Ref. [42] will be given in a forthcoming article.

We briefly mention that the injection chain is composed of a HPLC pump, and a 30μm glass capillary facing a catcher used to extract the liquid from the vacuum chamber, both placed in a differentially pumped tube. The liquid jet is set perpendicular to the photon beam propagation axis and to the spectrometer lens axis. We performed measurements on 0.5M and 1M KCl aqueous solutions that were prepared by mixing >99% KCl salt with deionized water. Filtering and degassing procedures were systematically performed on the solutions. The pressure in the main chamber was kept below the $10^{-5}$ mbar range whereas it was kept at about $10^{-4}$ mbar in the differentially pumped tube when the liquid is injected. A spectrometer resolution of about 0.6eV is achieved with 500eV pass energy and 0.5mm slits. The photon energy resolution achieved at 5keV is about 0.6eV.

Computational: In the present work the microsolvated $K^+(H_2O)$ and $K^+(H_2O)_2$ clusters were considered with the optimized geometries taken from ref.[18]. The energies of the 1s single- and $2p^{-2}$ double-core hole states were computed by an occupation restricted multiple active space (ORMAS) single- and double-excitation configuration interaction (SDCI) method as implemented in the GAMESS-US package [43]. All core orbitals except for the considered ones were kept frozen in the computations. We used the 6-311+G* basis set for potassium and the 6-311G* one for oxygen and hydrogen atoms [44]

**Acknowledgments**

Experiments were performed at the GALAXIES beamline, SOLEIL Synchrotron, France (Proposal No. 20140160). The authors are grateful to the SOLEIL staff for assistance during the beamtime. N.V.K. thanks the Deutsche Forschungsgemeinschaft for financial support and Dr. V. Stumpf for help with calculations. We are grateful to Dr. M. Tashiro for valuable discussions. Campus France and the PHC SIAM exchange program are acknowledged for financial support.

**References**

[1] S. Eberhard, G. Finazzi, and F.A. Wollmann, Annu. Rev. Genet. **42**, 463 (2008).

[2] M. Cordes and B. Giese, Chem. Soc. Rev. **38**, 892 (2009).

[3] H. Ikeura-Sekiguchi and T. Sekiguchi Phys. Rev. Lett. **99**, 228102 (1991).

[4] M. A. Loi and A. Troisi, Phys. Chem. Chem. Phys., **16**, 20277 (2014).

[5] A. Nilsson, H. Tillborg, and N. Mårtensson, Phys. Rev. Lett. **67**, 1015 (1991).

[6] N. V. Dobrodey, L. S. Cederbaum, and F. Tarantelli, Surf. Sci. **402-404**, 508 (1998).

[7] R. P. Vasquez, C. U. Jung, M.-S. Park, H.-J. Kim, J. Y. Kim, and S.-I. Lee, Phys. Rev. B **64**, 052510 (2001).

[8] N. V. Dobrodey, A. I. Streltsov, L. S. Cederbaum, C. Villani, and F. Tarantelli, Phys. Rev. B **66**, 165103 (2002)

[9] N. V. Dobrodey, A. I. Streltsov, and L. S. Cederbaum, Phys. Rev. A, **65**, 023203 (2002).

[10] A. I. Streltsov, N. V. Dobrodey, and L. S. Cederbaum, J. Chem. Phys. **119**, 3051 (2003).

[11] N. V. Kryzhevoi, N. V. Dobrodey, and L. S. Cederbaum, J. Chem. Phys. **122**, 104304 (2005)

[12] N. V. Kryzhevoi and L. S. Cederbaum, J. Chem. Phys. **123**, 154308 (2005).

[13] N. V. Kryzhevoi and Cederbaum L. S. J. Chem. Phys. **130**, 084302 (2009).

[14] E. F. Aziz, M. H. Rittmann-Frank, K. M. Lange, S. Bonhommeau, and M. Chergui, Nature Chemistry **2**, 853 (2010)

[15] O. Björneholm, A. Nilsson, A. Sandell, B. Hernnäs, and N. Mårtensson, Phys. Rev. Lett. **68**, 1892 (1992).

[16] A. Föhlisch, P. Feulner, F. Hennies, A. Fink, D. Menzel, D. Sanchez-Portal, P. M. Echenique, and W. Wurth, Nature **436**, 373-376 (2005)

[17] F. Gel'mukhanov and H. Ågren, Phys. Rev. A, **54**, 3960 (1996).

[18] W. Pokapanich, N. V. Kryzhevoi, N. Ottosson, S. Svensson, L. S. Cederbaum, G. Öhrwall, and O. Björneholm, J. Am. Chem. Soc. **133**, 13430 (2011)

[19] L. S. Cederbaum, J. Zobeley, and F. Tarantelli, Phys. Rev. Lett. **79**, 4778 (1997).

[20] P. Slavíček, B. Winter, L. S. Cederbaum, and N. V. Kryzhevoi, J. Am. Chem. Soc. **136**, 18170 (2014).

[21] J. Zobeley, R. Santra, and L. S. Cederbaum, J. Chem. Phys. **115**, 5076 (2001).

[22] I. Unger, R. Seidel, S. Thürmer, M. N. Pohl, E. F. Aziz, L. S. Cederbaum, E. Muchová, P. Slavíček, B. Winter, and N. V. Kryzhevoi, Nature Chem. (2017) doi:10.1038/nchem.2727.

[23] K. Ueda, private communication.

[24] S. Nishikida and S. Ikeda, J. Electron Spectrosc. Relat. Phenom., **13**, 49 (1978).

[25] W. Pokapanich, H. Bergersen, I. L. Bradeanu, R. R. T. Marinho, A. Lindblad, S. Legendre, A. Rosso, S. Svensson,

O. Björneholm, M. Tchaplyguine, G. Öhrwall, N. V. Kryzhevoi, and L. S. Cederbaum, J. Am. Chem. Soc., **131**, 7264 (2009).

[26] B. Winter and M. Faubel, Chem. Rev. **106**, 1176 (2006).

[27] M. O. Krause and J. H. Oliver J. Phys. Chem. Ref. Data **8**, 329 (1979).

[28] C. Fang, W.-F Li, R.S. Koster, J. Klimes, A. Van Blaaderen, and M. A. Van Huis, Phys. Chem. Chem. Phys., **17**, 365 (2015).

[29] B. Cleff and W. Mehlhorn, Z. Physik **219**, 311 (1969).

[30] K. Siegbahn, C. Nordling, A. Fahlman, R. Nordberg, K. Hamrin, J. Hedman, G. Johansson, T. Bergmark, S.-E. Karlsson, I. Lindgren, and B. Linderg, Nova Acta Regiae Soc. Sci. Upsaliensis Ser. IV, 20 (1967).

[31] J. Vayrynen, R. N. Sodhi, and R. G. Cavell, J. Chem. Phys. **79**, 5329 (1983).

[32] E. J. Aitken, M. K. Bahl, K. D. Bomben, J. K. Gimzewski, C. S. Nolan, and T. D. Thomas, J. Am. Chem. Soc., **102**, 4873 (1980).

[33] B. Lohmann and S. Fritzsche, J. Phys. B. **27**, 2919 (1994).

[34] M.J. Blandamer and M.F. Fox, Chem. Rev. **70**, 59 (1970).

[35] N. V. Kryzhevoi, F. Tarantelli, and L. S. Cederbaum, Chem. Phys. Lett. **626**, 85 (2015).

[36] B. Winter, R. Weber, W. Widdra, M. Dittmar, M. Faubel, and I. V. Hertel, J. Phys. Chem. A **108**, 2625 (2004).

[37] H. Siegbahn, M. Lundholm, S. Holmberg, and M. Arbman, Phys. Scr. **27**, 431 (1983).

[38] C. H. H. Van Deurzen, J. G. Conway, and S. P. Davis, J. Opt. Soc. Am. **63**, 158 (1973)

[39] Y. Markus, Chem. Rev. **109**, 1346 (2009).

[40] D. Céolin, J.M. Ablett, D. Prieur, T. Moreno, J.P. Rueff, T. Marchenko, L. Journel, R. Guillemin, B. Pilette, T. Marin, and M. Simon, Journal of Electron Spectroscopy and Related Phenomena, **190**, 188192 (2013),

[41] J. P. Rueff, J.M. Ablett, D. Céolin, D. Prieur, T. Moreno, V. Balédent, B. Lassalle-Kaiser, J. E. Rault, M. Simon, and A. Shukla, Journal of Synchrotron Radiation, **22**, 175 (2015).

[42] M. Faubel, S. Schlemmer, and J. P. Toennies, Z. Phys.D **10**, 269 (1988).

[43] M. W. Schmidt, K. K. Baldridge, J. A. Boatz, S. T. Elbert, M. S. Gordon, J. H. Jensen, S. Koseki, N. Matsunaga, K. A. Nguyen, S. J. Su, T. L. Windus, M. Dupuis, and J. A. Montgomery, J. Comput. Chem. **14**, 1347 (1993).

[44] J.-P. Blaudeau, M. P. McGrath, L. A. Curtiss, and L. Radom, J. Chem. Phys. **107**, 5016, (1997).